\documentstyle[aps,epsfig]{revtex}
 
\tighten
\begin{document}

\title{\bf Reheating in the Presence of Noise}

\author{V. Zanchin$^{1,3}$\footnote[1]{zanchin@het.brown.edu}, 
A. Maia Jr.$^{1,2}$\footnote[2]{maia@ime.unicamp.br},  
W. Craig$^{4}$\footnote[3]{craigw@math.brown.edu} and R. Brandenberger$^{1}$\footnote[4]{rhb@het.brown.edu}} 
  
\smallskip
\address{~\\$^1$Physics Department, Brown University, 
Providence, RI 02912, USA.}
 
\address{~\\$^2$Departamento de Matem\'atica Aplicada, 
     Universidade Estadual de Campinas, 
     13083 - 970, Campinas, SP, Brazil.}

\address{~\\$^3$Departamento de F\'{\i}sica, Universidade 
Federal de Santa Maria, \\97119-900, Santa Maria, RS, Brazil.}

\address{~\\$^4$Mathematics Department and Lefschetz Center for 
Dynamical Systems, Brown University, Providence, RI 02912, USA.}

\maketitle

\vskip 1.5cm
\begin{abstract}

\noindent 
Explosive particle production due to parametric resonance is a crucial 
feature of reheating in inflationary cosmology. Coherent oscillations 
of the inflaton field act as a periodically varying mass in the evolution 
equation for matter fields which couple to the inflaton. This in turn 
results in the parametric resonance instability. Thermal and quantum 
noise will lead to a nonperiodic perturbation in the mass. We study 
the resulting equation for the evolution of matter fields and demonstrate 
that noise (at least if it is temporally uncorrelated) will increase the 
rate of particle production. We also estimate the limits on the magnitude 
of the noise for which the resonant behavior is qualitatively unchanged.
\end{abstract}
 
\vfill

\setcounter{page}{0}
\thispagestyle{empty}

\vfill

\noindent BROWN-HET-1072\hfill      August   1997.

\noindent hep-ph/yymmdd \hfill Typeset in REV\TeX

\vfill\eject

\baselineskip 24pt plus 2pt minus 2pt

\section{Introduction}

Over the past few years, it has been realized that explosive particle 
production at the end of the period of inflation is a crucial aspect 
of inflationary cosmology. At the end of the period of exponential 
expansion, the energy density of matter and radiation is exponentially 
small, and without a very fast transfer of energy from the inflaton to 
ordinary matter, it is not possible to obtain a high post-inflationary 
temperature. 

As was first pointed out in \cite{TB} and discussed in more detail in 
\cite{kls94,stb,kls97} and many other recent papers$^{\cite{recent}}$, an oscillating scalar 
field induces a parametric resonance instability in the mode equations 
of any bosonic matter fields which couple to it. In particular, this 
applies to the inflaton, the scalar field responsible for inflation. 
At the end of the period of exponential expansion of the Universe, the 
inflaton is predicted to be performing homogeneous oscillations about 
its vacuum state. This will lead to instabilities in the mode equations 
of any bosonic matter field which couples to the inflaton, and this 
instability corresponds to explosive particle production.

Instabilities, due to resonance effects,
are in general quite sensitive to the presence or absence 
of noise. In a real physical system we expect some non-periodic noise 
in the evolution of the inflaton. Such noise could be due to quantum 
fluctuations (the same quantum zero point oscillations which are believed 
to be the source of classical density perturbations$^{\cite{flucts}}$ in 
the inflationary Universe scenario) or thermal fluctuations.
\footnote{Note that in the usual analysis, the inflaton is treated as 
a classical background field.} In both cases, the amplitude of this noise 
is expected to be very small. Nevertheless, it is important to analyse the 
effects of this noise on parametric resonance.

In this paper, we study the effects of noise in the inflaton field on 
the evolution equation of matter fields which couple to the inflaton. 
As a first step, we take the noise to be homogeneous in space. In a 
subsequent paper we plan to study the (more realistic) case of 
inhomogeneous noise. Our main result is that noise which is not 
correlated temporally will on the average lead to an increase in 
the rate of particle production. We also derive limits on the 
amplitude of the noise for which the parametric resonance behavior 
persists. This result eliminates a further doubt on the effectiveness 
of resonance in realistic inflationary models. Most of our results 
apply for both narrow and broad-band resonance. For simplicity we 
neglect the expansion of the Universe; however, we do not believe 
that including effects of expansion would change our main conclusions.   
  
  In this paper we shall consider a simple model of reheating 
in the presence of a spatially homogeneous noise term. The inflaton 
field is taken to be a real scalar field and is denoted by $\varphi$. 
In the period immediately after inflation, $\varphi(t)$ is assumed to 
be oscillating coherently about (one of) its ground state(s). We shall, 
however, include a small aperiodic perturbation, i.e.
\begin{equation}
    \varphi(t) = A cos(\omega t) + q(t) \, ,
\label{noise}
\end{equation}
where $\omega$ is the natural frequency of oscillation of $\varphi$ 
and $q(t)$ denotes the noise.

We shall (following \cite{TB,kls94,stb}) take $\varphi$ to be coupled to 
a second scalar field $\chi$ (which represents matter) via an 
interaction Lagrangian which is quadratic in $\chi$, for example 
of the form
\begin{equation}
     {\cal L}_{int} \propto {1 \over 2} \varphi \chi^2 \, .
\end{equation}
To simplify the discussion, we neglect nonlinearities in the equation 
of motion for $\chi$. Such nonlinearities may very well be important 
and lead to an early termination of parametric resonance. This issue 
has recently been discussed extensively, see e.g. \cite{ac96,thermal,tp96}, 
but is not the topic of our paper. Instead, we are interested whether the 
presence of noise such as included in (\ref{noise}) will effect the onset 
of resonance.

For simplicity, we will neglect the expansion of the Universe. In 
\cite{TB,kls94,stb} it has been shown that the expansion can be 
included without difficulty, and that it does not prevent the onset 
of the parametric resonance instability. Since the equation of motion 
for $\chi(x,t)$ is linear and translation invariant, the Fourier modes 
evolve independently. We can immediately write down the evolution 
equations for the Fourier modes of $\chi$, denoted by $\chi_k$, in 
the case of ``coherent" noise. They take the form 
\begin{equation}
     \ddot {\chi_k} + \left[\omega_k^2 + (p(\omega t) 
        + q(t))\right] \chi_k = 0 \, ,
\label{eq:n1}
\end{equation}
where $p(x)$ is a periodic function whose frequency is $2 \pi$,  
$q(t)$ represents the noise which we consider as a perturbation 
of the driving function $p(t)$, and 
\begin{equation}
    \omega_k^2 = m_{\chi}^2 +  k^2 \, .  
\end{equation}
Note that our analysis applies both to the case of 
narrow-band$^{\cite{TB,stb}}$ ($p / {\omega^2} \ll 1$) and 
broad-band$^{\cite{kls94}}$ ($p / {\omega^2} \gg 1$) resonance.

By assumption, $q(t)$ is a small perturbation of $\varphi(t)$. Hence, 
it is suppressed compared to $p(t)$, and in order to represent this, 
we write
\begin{equation}
     q(t) = g \omega^2 n(t) \, , \label{noise2}
\end{equation}
where $n(t)$ is a dimensionless ``noise function" of amplitude 1, 
and where we have introduced the dimensionless small coupling $g$ 
by extracting the dimensions of $q$ by means of the factor $\omega^2$.

We wish to consider how the presence of noise in the inflaton effects 
the excitation of the matter field components $\chi_k$ from their vacuum 
state. Another way in which thermal noise can effect the calculation has 
recently been studied in \cite{hotta} where it was shown that parametric 
resonance is also effective if we consider the field $\chi$ to be initially 
in a thermal state.

We can study this problem in three ways (see e.g. \cite{RB97} for a 
recent review). First, we study the effects of the noise on the 
solutions of the classical equation of motion (\ref{eq:n1}) using 
the method of successive approximations and the Furstenberg theorem
on products of random matrices. Next, we use the Bogoliubov mode 
mixing technique $^{\cite{bogol}}$ to analyze the change in the 
solutions due to the noise. This method is closely related to a 
consistent semiclassical analysis which treats the excitation of 
the $\chi$ field as a problem of quantum field theory in a classical 
background inflaton field (see e.g. Appendix B of \cite{stb}). Finally, 
we study the effects of the noise using the Born approximation and 
compare with numerical results.

\section{Exact Results}

The principal results of this section are that the exponential growth
rate of solutions is a continuous quantity with respect to $q(t)$, 
so that a small addition of noise will not change it overly. 
Furthermore, under reasonable assumptions of decorrelation of 
the noise, the growth rate is shown to be always strictly increased 
by the noise. For these considerations it proves convenient to 
rephrase our basic second order differential equation (\ref{eq:n1}) 
as a first order $2\times 2$ matrix differential equation
\begin{equation} \label{matrixeq}
   {\dot \Phi_q} \, = \, M(q(t), t) \Phi_q 
\end{equation}
with initial conditions $\Phi_q(0,0) = I$. Here, $M(q(t), t)$ is 
the matrix
\begin{equation} \label{matrix}
    M(q(t), t) \, = \, 
      \left( \begin{array}{cc}
      0 & 1  \\  
      - (\omega_k^2 + p(t) + q(t)) & 0 
       \end{array} \right) \, ,
\end{equation}
and $\Phi_q(\tau,\sigma)$ is the fundamental solution (or transfer) matrix
from time $\sigma$ to time $\tau$; 
\begin{equation} \label{matrixsol}
     \Phi_q(t, 0) \, = \, \left( \begin{array}{cc} 
                          \phi_1(t;q) & \phi_2(t;q) \\
                                   {\dot\phi_1(t;q)} & {\dot\phi_2(t;q)}
                         \end{array} \right) \, ,
\end{equation}
consisting of two independent solutions $\phi_1(t;q)$ and $\phi_2(t;q)$ 
of the second order equation (\ref{eq:n1}).

For vanishing noise, i.e. $q(t) = 0$, the content of Floquet theory is
that the solution of (\ref{matrixeq}) can be written in the form
\begin{equation} \label{tm0matrix}
     \Phi_0(t, 0) \, = \, P_0(t) e^{Ct} \, ,
\end{equation}
where $P_0(t)$ is a periodic matrix function with period $T = 1/\omega$, 
and $C$ is a constant matrix whose spectrum in a resonance region is 
$spec(C) = \{\pm \mu(0)\}$.

We would like the noise $q(t)$ to correspond to random quantum or thermal 
fluctuations superimposed on the periodic classical oscillation of the 
inflaton field $\varphi(t)$. Our picture is that of fluctuations driven 
by Brownian motion, however for the purposes of deriving the exact results 
to be presented here, it is sufficient to make certain statistical 
assumptions about the noise $q(t)$. These are phrased in terms of a 
sample space $\Omega$ from which the realizations $\kappa$ of the 
noise $q(t) = q(t; \kappa)$ are drawn. On the sample space $\Omega$
there is a probability measure $dP(\kappa)$, and expectation values 
of functions $f(\kappa)$ with respect to this measure are denoted by 
\[
   E(f) \, =  \int dP(\kappa) f(\kappa) \, .
\]
For our purposes we may take $\Omega = C({\mathbf R})$, the space
of bounded continuous functions on ${\mathbf R}$, and $dP(\kappa)$
on $\Omega$ a translation invariant measure. 
We assume that the noise is ergodic, which is to say that
\begin{equation} \label{ergodic}
     E(f(q)) \, = \, \lim_{t \rightarrow \infty} {1 \over t} 
               \int_0^t d\tau \ f(q( \cdot + \tau; \kappa_0))
\end{equation}
for almost all $dP(\kappa)$ realizations $\kappa_0$ of the noise.

The first relevant mathematical result (Proposition 1, see Appendix) 
is that the growth rate (the generalized Floquet  exponent, or Lyapunov 
exponent) of the solutions of (\ref{matrixeq}) is well defined by the limit
\begin{equation} \label{growth}
       \mu(q) \, = \, \lim_{N \rightarrow \infty} {1 \over {N T}} 
          \log \| \Pi_{j=1}^N \Phi(jT,(j-1)T) \| \, ,
\end{equation}
where $\| \cdot \|$ denotes some matrix norm (the dependence on the 
specific norm drops out in the large $N$ limit). The limit in (\ref{growth}) 
exists for almost every sample (with respect to $dP(\kappa)$), and it
is almost everywhere $dP(\kappa)$ constant, which is to say that 
it depends only upon the the statistics of samples $q(t;\kappa)$ and 
not on the individual realizations.

The first qualitative result (Proposition 2, see Appendix) is that the 
growth rate $\mu(q)$ is continuous in $q$ in an appropriate topology 
on $\Omega$ (the topology on $C({\mathbf R})$ of uniform convergence 
on compact sets). 
In particular, this implies that if we write the noise as in (\ref{noise2}) 
in terms of a dimensionless strength $g$ and view the statistics of the 
noise as a measure on the function space of $n(t)$, then $\mu(q)$ converges 
to $\mu(0)$ as $g$ converges to zero. We remark that Propositions 1 and 2
do not depend on the fact that the system is one-dimensional, and will thus 
also hold for inhomogeneous noise. 

The limit (\ref{growth}) also describes $\mu(0)$, which is implicitly 
a function of the background periodic potential $p(t)$; it shows 
in particular that $\mu(0) \geq 0$. A second qualitative estimate
(Theorem 3, see Appendix and \cite{kotani}) is that, if the support 
of the probability measure $dP(\kappa)$ on the space $C({\mathbf R})$
includes the sample $q(t) = 0$, then $\mu(q) > 0$ whenever $\mu(0) > 0$.

%In the following sections we will 
%give quantitative estimates for the magnitude of $\mu(q) - \mu(0)$. 
%
%For the moment, however, we will concentrate on general results. The 
%following result (Proposition 3, Appendix) is special to one-dimensional 
%second order systems of ordinary differential equations and will hence 
%not necessarily hold for inhomogeneous noise. Given the probability 
%space $\Omega = C({\Re})$ of functions $p(t) + q(t)$, then the growth 
%rate $\mu(p + q)$ defined in (\ref{growth}) (viewed for a moment without 
%the averaging) depends only of the statistics of the probability measure 
%on $\Omega$ and not on the individual sample $\kappa$. In addition if the 
%support of the probability measure includes $q(t) = 0$, then 
%$\mu(q + p) > 0$ whenever $\mu(p) > 0$.

The results cited above do not give insight into the quantitative value 
of the growth rate in the presence of noise. A first step in this direction 
is given by the following theorem. Suppose that 
the statistics of the noise are such that
\begin{description} 

\item{(i)} The noise $q(t)$ is uncorrelated in time on scales larger than $T$,
that is, $\{q(t; \kappa) \, : \, jT \leq t \leq (j + 1)T\}$ is independent of 
$\{q(t; \kappa) \, : \, lT \leq t \leq (l + 1)T\}$ for integers $l \neq j$, 
and is identically distributed.

\item{(ii)} Restricting the noise $q(t; \kappa)$ to the time interval 
$0 \leq t < T$, the samples $\{q(t; \kappa): 0 \leq t < T\}$ within the 
support of the probability measure fill a neighborhood, in $C(0,T)$, of the 
origin.
\end{description}

Hypothesis (i) implies that the noise is ergodic, and therefore the 
generalized Floquet exponent is well defined. We will show (Theorem 4, 
Appendix)  that in fact $\mu(q)$ is strictly larger than $\mu(0)$;
\begin{equation} \label{main}
     \mu(q) \, > \, \mu(0) \, ,
\end{equation}
which demonstrates that the presence of noise leads to a strict increase 
in the rate of particle production. This is a quantitative lower bound on 
the generalized Floquet exponent with noise. Our result is based on  an
application of Furstenberg's theorem, concerning the Lyapunov exponent of 
products of independent identically distributed random matrices 
$\{\Psi_j : j=1, ... ,N\}$. It states that (modulo certain assumptions 
which are shown to hold in the Appendix)
\begin{equation} \label{fu1}
     \lim_{N \rightarrow \infty} {1 \over {NT}} 
         \log \| \Pi_{j=1}^N \Psi_j \| \, = \,  \lambda \, > \, 0 \, , 
\end{equation}
where $\lambda = \lambda(q)$ depends again only on the statistics of 
$\Psi_j$,  and not on the individual samples. A further result is that 
for any nonzero vectors $v_1$ and $v_2$ and for almost all 
$\{ \Psi_j(\kappa) \}_{j=1}^\infty$
\begin{equation} \label{fu2}
   \lim_{N \rightarrow \infty} {1 \over {NT}} 
      \log \langle v_1, \Pi_{j=1}^N \Psi_j v_2 \rangle \, = \,  \lambda \, .
\end{equation}

In order to apply Furstenberg's theorem to obtain (\ref{main}), we start 
by factoring out from the transfer matrix $\Phi_q(t, 0)$ the contribution 
due to the evolution without noise;
\begin{equation} \label{factoriz}
     \Phi_q(t, 0) \, = \, \Phi_0(t, 0) \Psi_q(t, 0) \, = \, P_0(t) e^{Ct} 
     \Psi_q(t, 0) \, .
\end{equation}
The reduced transfer matrix $\Psi_q(t, 0)$ satisfies the following equation
\begin{equation} \label{Sdef}
     {\dot \Psi_q} \, = \, S(t; \kappa) \Psi_q \, = \, \Phi_0^{-1}(t)
     \left( \begin{array}{cc} 0 & 0 \\
             -q(t; \kappa) & 0
       \end{array} \right)\Phi_0(t) \Psi_q \, ,
\end{equation}
which can be written as a matrix integral equation
\begin{equation} \label{redint}
     \Psi_q(t, 0) \, = \, 
         I + \int_0^t d\tau \ S(\tau; \kappa) \Psi_q(\tau, 0) \, .
\end{equation}
Solutions $\Psi_q(t, 0)$ to (\ref{redint}) are constructed in the next 
section, using the method of successive approximation, giving rise to 
the transfer matrices $\Psi_q(jT, (j-1)T)$ which are the fundamental 
solution matrices for the period intervals of the background potential.
By properties (ii) of decorrelation of the noise, the quantities
$\Psi_q(jT, (j-1)T)$ are independent and identically distributed 
for different integers $j$, and we can  apply the Furstenberg theorem 
to the following decomposition of $\Psi_q(NT, 0)$;
\[
     \Psi_q(NT, 0) \, = \, \Pi_{j=1}^N \Psi_q(jT, (j-1)T) \, .
\]
We may choose for instance the vector $v_1$ in (\ref{fu2}) to be an 
eigenvector of $\Phi_0(T,0)^t = (P_0(T) e^{CT})^t$, the transpose of the 
transfer matrix of the system without noise, with eigenvalue $e^{\mu(0)T}$. 
Then (\ref{fu2}) becomes 
\begin{eqnarray}
  {1 \over {NT}} \log |\langle v_1, \Phi_q(NT, 0) v_2 \rangle | 
  & = & {1 \over {NT}} \log 
        |\langle v_1, P_0(NT) e^{CNT} \Psi_q(NT, 0) v_2 \rangle | 
  \nonumber \\
  & = & {1 \over {NT}} \log ( e^{\mu(0)NT} 
        |\langle v_1, \Psi_q(NT, 0) v_2 \rangle |) \\
  & = & \mu(0) \, + \, {1 \over {NT}} \log 
        |\langle v_1, \Pi_{j=1}^N \Psi_q(jT, (j-1)T) v_2 \rangle | \, .
  \nonumber
\end{eqnarray}
Taking the limit $N \rightarrow \infty$ and applying (\ref{fu1}) 
and (\ref{fu2}) we obtain
\begin{equation}
   \mu(q) \, = \, \mu(0) + \lambda \, > \, \mu(0)
\end{equation}
which proves the main result (\ref{main}). Note in particular that 
(\ref{main}) implies that even for modes which without noise are in 
a stability band ($\mu(0) = 0$) there will be exponential particle 
production in the presence of noise. The fact that there is particle 
production due to the presence of noise is a nontrivial result, and the 
fact that the rate of particle production is exponential may be 
understood physically as a sign of stimulated emission. Note that, strictly speaking, the ergodic hypothesis is only satisfied in the 
$t \rightarrow \infty$ limit. Finite time intervals, such as the characteristic time of reheating $\mu^{-1}$, may be insufficient for the decorrelation of the noise. In this case, we may observe transient effects (see Section VII).

So far, our results have not given any quantitative upper bound for the 
growth rate of modes of $\chi$ in the presence of noise. Additionally,
they do not show how the evolution of an individual solution is modified,
as compared to what occurs in the system without noise. In the following 
sections we will give estimates on the magnitude of noise for which we 
can demonstrate that the parametric resonance behavior remains qualitatively 
unchanged. These estimates are then used in the proof in the appendix of 
the main result (\ref{main}).

\section{An Estimate Using Successive Approximations}

In this section we sketch the construction of the fundamental 
solution matrix $\Psi_q(t,0)$ for system (\ref{Sdef}), showing
in particular that $\{ \Psi_q(jT,(j-1)T) \}_{j=1}^\infty$ represents 
a sequence of independent identically distributed random matrices. 
The starting point for this estimate is the integral equation 
(\ref{redint}) for the reduced transfer matrix $\Psi(t, 0)$ (dropping 
the subscript $q$ for notational ease). We solve this equation by 
constructing a sequence of approximate solutions 
\begin{equation}
       \Psi_{n+1}(t,0) \, = \, I \, + \, 
         \int_0^t d\tau \ S(\tau; q(\tau)) \Psi_n(\tau, 0) 
\end{equation}
with $\Psi_0 = I$. The differences between successive terms of the 
sequence satisfy
\begin{equation}
   \Psi_{n+1}(t, 0) - \Psi_n(t, 0) \, = \, \int_0^t d\tau \ S(\tau; q(\tau)) 
       \bigl( \Psi_n(\tau, 0) - \Psi_{n-1}(\tau, 0) \bigr) \, .
\end{equation}
We will apply this equation for evolution over the time period $T$.

By induction is can easily be shown that the successive differences 
satisfy the estimate
\begin{equation}\label{bigpsi}
   \| \Psi_{n+1}(T, 0) - \Psi_n(T, 0) \| \, \leq \, M^{n+1} 
    {{T^{n+1}} \over {(n+1)!}} \, ,
\end{equation}
where the constant $M$ is
\begin{equation}\label{constM}
      M \, = \, \sup_{0 \leq s \leq T} \|S(s; q)\| \, .
\end{equation}
{}From the definition of the matrix $S$ (see (\ref{Sdef}) and 
(\ref{tm0matrix})) it follows that (if the norm $\| \cdot \|$ is 
taken to be the supremum norm or the norm mentioned after (\ref{growth}))
\begin{equation}
M \, \leq \, e^{2 \mu(0) T} \|q(\cdot; \kappa)\| \, .
\end{equation}

The following telescoping sum then describes the fundamental 
solution matrix;
\begin{equation}
   \Psi(t,0) \, = \, \sum_{n=1}^\infty \bigl(\Psi_n(t,0) - \Psi_{n-1}(t,0)\bigr) + I ~,
\end{equation}
and by using (\ref{bigpsi}) we estimate that
\begin{eqnarray}
   \| \Psi(t,0) - I \| & \, \leq \, & \sum_{n=1}^\infty {(Mt)^n \over n\!} 
   \nonumber  \\
    & \, = \, & e^{Mt} -1  \, \leq \, (Mt)  e^{Mt} ~.   
\end{eqnarray}
The fundamental solution of (\ref{matrixeq}) is 
$\Phi_q(t,0) = \Phi_0(t,0)\Psi_q(t,0)$, therefore the deviation between
the solution with noise and the solution without it can be measured
by the estimate
\begin{eqnarray}
   \| \Phi_q(t,0) - \Phi_0(t,0) \| & \, = \, & 
   \| \Phi_0(t,0)(\Psi_q(t,0) - I) \| \\
   & \, \leq \, & \|\Phi_0(t,0) \| \| R_q(t,0) \|  ~, \nonumber
\end{eqnarray}
where $R_q(t,0) = (\Psi_q(t,0) - I)$. When we set 
$q(t;\kappa) = g \omega^2 n(t;\kappa)$, and $g$ is taken to zero, then
the constant $M$ in (\ref{constM}) also  converges to zero, therefore
for fixed time $t$ the error $\| R_q(t,0) \| \leq (Mt)\exp(Mt)$ is
small and thus solutions of the two equations which have the same 
initial data are shown to be close. 

We will now make use of the above estimate in order to bound from above 
the difference $\mu(q) - \mu(0)$. We consider evolution over a time 
interval $NT$ and break up the transfer matrix into matrices 
corresponding to single periods $T$. From (\ref{factoriz}) and 
(\ref{growth}) we obtain
\begin{equation} \label{differ}
   \mu(q) - \mu(0) \, = \, \lim_{N \rightarrow \infty} {1 \over {NT}} 
      E \bigl( \log \bigl[ {{\| \Pi_{j=1}^N (P_0(T) e^{TC} 
    \Psi_q(jT,(j-1)T)) \|} \over {\| \Pi_{j=1}^N e^{TC} \|}} \bigr] \bigr) \, .
\end{equation}
Making use of the periodicity of $P_0$, we can write the norm of the 
product in the numerator (denoted by $NUM$) as
\[
     NUM \, = \, \| \Pi_{j=1}^N \bigl( e^{TC} 
        + \Delta_q(jT, (j-1)T) \bigr) \| \, ,
\]
where $\Delta_q$ represents the difference in between the transfer 
matrices with and without noise. We need to bound its contribution 
in magnitude from above. Due to the fact that the noise is uncorrelated 
over time intervals larger than $T$, this needs to be done only for one 
period. Introducing the symbol
\begin{equation} \label{bigtheta}
    \Theta \, = \, E \bigl( {{\| \Delta_q(T, 0) \|} 
           \over {\| e^{TC} \|}} \bigr) \, ,
\end{equation}
we can bound the difference in the generalized Floquet exponents 
(\ref{differ}) by
\begin{equation} \label{differ2}
     \mu(q) - \mu(0) \, \leq \, \lim_{N \rightarrow \infty} {1 \over {NT}} 
        \log(1 + \Theta)^N \, = \, {1 \over T} \log(1 + \Theta) \, ,
\end{equation}
which for low amplitude noise can be approximated by $\Theta / T$.

Making use of the fact that $\mu(0) / \omega \ll 1$, it follows that for 
noise of the form given by (\ref{noise2}), $\Theta$ is of the order 
$g \omega^2 T^2$.\footnote{One factor of $T$ comes from the time integration in the transfer matrix, the second arises since a factor of $T$ must be inserted in the matrix element in (\ref{Sdef}) if both components of $\Psi$ are to have the same dimensions.} Hence, from (\ref{differ2}) we conclude that if
\begin{equation} \label{estimate1}
       g \, \ll \, {{\mu(0)} \over {\omega}}
\end{equation}
the effects of the noise do not significantly effect parametric resonance.
Furthermore, the estimate (\ref{bigpsi}) shows that solutions of (6) 
cannot deviate by more than $MT\exp(MT)$ from $\Phi_0(t,0)$, over
time intervals of length $T$. 

\section{A Second Estimate Using Successive Approximations}
 
The growth rate $\mu(q)$ of the previous section is the result of a
limiting process over many periods $T$ of the background periodic 
excitation $p(t)$. It may be, however, that the noise acts over small 
characteristic subintervals of length $\Delta t$, and the reheating 
epoch only encompasses a finite and relatively small number of periods
of the background. By studying directly the second order differential 
equation (3) with an estimate based on the method of successive 
approximations, we will exhibit in such a situation a bound in terms 
of the coupling constant $g$ on the deviation of solutions perturbed 
by the presence of noise from the unperturbed solution. In case this 
deviation is small when compared with the size of the unperturbed 
solution, we can conclude that the growth due to parametric resonance 
is not destroyed by the presence of the noise term. We obtain an estimate 
on $g$ for which this holds; it turns out to be of the same character 
as estimate (\ref{estimate1}).

The method is to rewrite equation (\ref{eq:n1}) as an integral equation and 
to determine the general solution in terms of a series of successive 
approximations. In the following we use a well-known result for 
differential equations. Let $\chi(t)$ be the solution of the differential 
equation (we drop the index $k$ on $\chi$ to simplify the notation)
\begin{equation}
     \ddot {\chi} + \left[\omega_k^2 + (p(t) + q(t))\right] \chi = 0 \, ,
\label{eq:a1}
\end{equation}
subject to given initial conditions, and let $\chi_h$ be the solution of 
the unperturbed equation
\begin{equation}
     \ddot {\chi} + \left[\omega_k^2 + p(t)\right] \chi = 0 \, ,
\label{eq:a2}
\end{equation}
satisfying the same initial conditions. Then, Eq. (\ref{eq:a1}) can be 
rewritten as an inhomogeneous Volterra equation$^{\cite{Volterra}}$
of the second kind
\begin{equation}
    \chi(t) = \chi_h(t) + \int_{t_i}^{t}\left[{\phi_1(t^{'})\phi_2(t) 
    - \phi_1(t)\phi_2(t^{'})\over W}\right]\left[-q(t^{'})\chi(t^{'})
    \right] dt^{'} \, , 
\label{eq:a3}
\end{equation}
where $\phi_1$ and $\phi_2$ are two independent solutions of equation
(\ref{eq:a2}) and $W$ is their Wronskian, which is independent of time 
(by Abell's formula, see e.g. \cite{Rabenstein}).  We then apply the 
method of successive approximations starting with the
unperturbed solution $\chi_h(t)$. We define recursively 
\begin{eqnarray}
   \chi_0(t) &=&\chi_h(t) \, ,     \nonumber\\
   \chi_{n}(t) &=&\chi_h(t) + \int_{t_i}^{t}\left[{\phi_1(t^{'})\phi_2(t) 
      - \phi_1(t)\phi_2(t^{'})\over W}\right]\left[-q(t^{'})\chi_{n-1}(t^{'})
        \right] dt^{'} \, \quad n \geq 1 \, .
\label{eq:a4}
\end{eqnarray}
{F}rom the above, we get an appropriate estimate for the successive 
approximations
\begin{equation}
     \chi_{n+1}(t)-\chi_n(t)= \int_{t_i}^{t}\left[{\phi_1(t^{'})\phi_2(t) 
     - \phi_1(t)\phi_2(t^{'})\over W}\right]\left[-q(t^{'})\right]
       \left[\chi_n(t^{'})-\chi_{n-1}(t^{'})\right] dt^{'} \, .
\label{eq:a5}
\end{equation}
The function $p(t)$ is periodic, so that the two independent solutions
of equation (\ref{eq:a2}) can be written in Floquet's form$^{\cite{Floquet}}$
\begin{equation}
     \phi_1(t)= e^{\mu(0) t} p_1(t)\, , \hspace*{1cm} 
     \phi_2(t)= e^{-\mu(0) t} p_2(t)\, ,
\label{eq:a6}
\end{equation}
where $p_1$ and $p_2$ are both periodic functions of time with period $T$,
and $\mu(0)$ is the Floquet exponent discussed in the previous section.
Let us focus on the behavior of the solution over an arbitrary time 
interval $I = [t_i, t_f]$, of length $\Delta T$. Define
\begin{equation}
     M_1 = \max_{0 \leq t \leq T} |p_1(t)| \, , \quad 
     M_2 = \max_{0 \leq t \leq T} |p_2(t)| \, .   \\     
\label{eq:a7}
\end{equation}
According to our assumptions, the perturbation $q(t)$ is bounded by 
$g \omega^2$, 
\begin{equation}
    \max_{ t \in {\bf R}} \left|q(t)\right| \leq g \omega^2 < +\infty \, .
\label{eq:a8} 
\end{equation}

Defining the norm
\begin{equation}
    \| \chi_n - \chi_{n-1}\| = \max_{t \in I}
    | \chi_n(t)-\chi_{n-1}(t) | \, , 
\label{eq:a9}
\end{equation}
and using equations (\ref{eq:a5})--(\ref{eq:a8}) it follows that 
for $t \in I$ then
\begin{equation} \label{eq:a10}
     | \chi_{n+1}(t) -\chi_n(t) | \leq \, 
     \Bigl( 2 g \omega^2 {{M_1 M_2} \over {|W|}} 
     e^{\mu(0)\Delta T} \Bigr)^{n+1} {(t - t_i)^{n+1} \over (n+1)! }
     \|\chi_h\| \, . 
\end{equation}
Define 
\begin{equation}
   \alpha = 2g \omega^2 {M_1 M_2 \over |W| } e^{\mu(0) \Delta T} \, .
\label{eq:a11}
\end{equation}

Let us now estimate the distance $d = \| \chi - \chi_h \|$. 
We consider the telescopic series
\begin{equation}
   \chi_{n+1}(t) - \chi_h(t) 
   = \sum_{j=0}^{n}(\chi_{j+1}-\chi_j) \, , 
\label{eq:a13}
\end{equation}
and use equations (\ref{eq:a10}) and (\ref{eq:a11}) and the triangle 
inequality to obtain
\begin{equation}
  | \chi_{n+1}(t) - \chi_h(t) | \, \leq \, \sum_{j=0}^{n} 
  | \chi_{j+1}(t)  - \chi_j(t) | \, \leq \, 
  \sum_{j=0}^{n} \alpha^{j+1} { (t - t_i)^{j+1} \over (j+1)! } 
  \| \chi_h \|  \, ,
\label{eq:a14}  
\end{equation}
where $\alpha$ is given by (\ref{eq:a11}). Taking the limit 
$n \to \infty$ and using the definition of the norm in (\ref{eq:a9})
we obtain 
\begin{equation}
  \| \chi-\chi_h \| \, \leq \, (e^{\alpha \Delta T} - 1 ) \| \chi_h \| 
   \, \leq \, \alpha \Delta T e^{\alpha \Delta T} \| \chi_h \| \, . 
\label{eq:a15}
\end{equation}
 
The last step is to ensure that the unperturbed solution has undergone 
appreciable growth by the end of the considered time interval.  Typical
initial data for the problem is of the form 
\begin{equation}
   \chi_h(t) \,  =  \, c_1 \phi_1(t) \,  +  \,  c_2 \phi_2(t) \,  ,
\label{eq:a16}
\end{equation}
with coefficients $c_1$ and $c_2$ which are of the order 1, therefore
$\|\chi_h\| = O(\exp (\mu(0) \Delta T))$. 

Note that the period of reheating in the absence of the noise is 
$N \mu(0)^{-1}$, with $N$ being a number of the order 
$2 \times {\rm log}(\eta / H)$ (where $H$ is the expansion rate 
of the Universe and $\eta$ is the energy scale of inflation). 
A typical value of $\eta / H$ is of the order $10^6$, and this 
implies that the number $N$ is not necessarily large. Hence, for 
practical purposes it is sufficient to control the effects of the 
noise over a time interval $\mu(0)^{-1}$. In addition, in order to 
prove that the effects of the noise are small compared to the 
unperturbed solution, we must consider time intervals over 
which the unperturbed solution has significant exponential 
growth, i.e. over which the growing mode solution of equation 
(\ref{eq:a2}), $\phi_1(t)$, dominates the decaying mode $\phi_2(t)$.  
This provides another reason to consider time intervals such that 
$\mu(0)\Delta T >1$.  To be definite we will take $\mu(0) \Delta T =2$. 

The maximal distance $d$ between $\chi$ and $\chi_h$  is controlled 
by (\ref{eq:a15}) over any interval $I$. If the amplitude of the noise 
is sufficiently small, then this distance will not be appreciable when
compared with the size of $\chi_h$. In order for this to be true, we require that
\begin{equation}
  \alpha \Delta T e^{\alpha \Delta T} \leq {1 \over 2} \, .
\label{eq:12}
\end{equation} 
The condition of smallness which emerges from (\ref{eq:12}) is that
\begin{equation} \label{eq:a12}
\alpha \Delta T = 2 e^2 g \omega^2 (M_1 M_2 / |W|) \Delta T << 1 \, .
\end{equation} 
To obtain an estimate of the order of magnitude of this constraint, 
we note that $|W| \, \sim \, M_1 M_2 \omega$. Inserting this into 
(\ref{eq:a12}) we obtain
\begin{equation}
  g \, \ll \, {\mu(0) \over {\omega}} \, , \label{eq:a18}
\end{equation}
which is the same requirement as was obtained in the previous section 
(\ref{estimate1}). Thus, in this case the parametric resonance growth 
of $\chi_h(t)$ is preserved in $\chi(t)$.  
     
Certainly there are better estimates that can be obtained if more information 
about the noise is provided. On the other hand, we have shown that for any
kind of small homogeneous noise, over time intervals of length 
$\Delta T = O(\mu(0)^{-1})$, parametric resonance is not destroyed by the 
presence of low amplitude noise in the inflaton field.

\section{An Analysis via the Bogoliubov Method}

Another approach which can be used to demonstrate that random noise with a small amplitude does not eliminate the parametric resonance instability is the Bogoliubov method$^{\cite{bogol}}$. This approach has the advantage of giving some information on how the
 noise effects the dynamics. However, it is less rigorous than the method of successive approximations and holds only in the case of
narrow resonance regime.
   
The evolution equation for $\chi_k(t)$ in the presence of the oscillating
inflaton field (see Eq. (\ref{eq:n1})) is (dropping the subscript $k$ and introducing an explicit expansion parameter $\varepsilon \ll 1$)
\begin{equation}
\ddot{\chi}(t) +\left[\omega_k^2 + \varepsilon\Bigl(p(\omega t) + q(t)
\Bigr)\right]\chi(t) = 0\, , \label{eq:v1}
\end{equation}
where $p(\omega t)$ is a function with period $2 \pi/\omega$ and $q(t)$ represents the
noise.  

{}For simplicity we consider a mode $k$ which (in the absence of noise) 
is in the first resonance band, i.e. for which $2 \omega_k = \omega - \Delta$,
where $\Delta$ is a small quantity. In this case we may write:
\begin{equation}
p(\omega t) = \omega_k^2\, \cos(2\omega_k+\Delta)t \, , \label{eq:v2}
\end{equation}
with $-\varepsilon\omega_k < 2\Delta < \varepsilon \omega_k$.
Following the Bogoliubov method \cite{bogol} we assume that, to first order
in $\varepsilon$, the solution of (\ref{eq:v1}) is of the form
\begin{equation}
\chi(t)= a(t)\cos(\omega_k+{\Delta\over2})t+
b(t)\sin(\omega_k +{\Delta\over2})t \,  ,\label{eq:v3}
\end{equation}
where $a(t)$ and $b(t)$ vary slowly in comparison with the trigonometric 
factors. Notice that the exact solution includes terms with frequencies higher
than $\omega_k+\Delta /2$. These terms are neglected in the first order 
solution since they are of higher order in $\varepsilon$.
   
   Inserting (\ref{eq:v3}) and (\ref{eq:v2}) in (\ref{eq:v1}) and keeping
only first order terms we obtain
\begin{eqnarray}
& &\left(\ddot{a}+ \omega\dot{b}-{\Delta\over2}\omega a +{\varepsilon\over2}\omega_k^2 a + \varepsilon q a\right)\cos(\omega_k +
 {\Delta\over2})t \nonumber \\ & &
+\left(\ddot{b}-\omega\dot{a}-{\Delta\over2}\omega b
 -{\varepsilon\over2}\omega_k^2 b  +\varepsilon q b\right)
      \sin(\omega_k +{\Delta\over2})t= 0\, . \label{eq:v4}
\end{eqnarray}
    
   It is clear that the coefficients of both the cosine and sine in the above
equation must be zero, which yields two differential equations for
the functions $a(t)$ and $b(t)$. We seek solutions of the form
\begin{equation}
a(t)= A(t)e^{\mu t}\, ,\quad\quad b(t) = B(t) e^{\mu t}\, , \label{eq:v5}
\end{equation}
where $\mu$ is a constant of order $\varepsilon$.  We then get
\begin{eqnarray} 
\ddot{A} + 2\mu\dot{A} +\omega \dot{B}+
\mu\omega B - {\Delta\over2}\omega A+{\varepsilon\over2}\omega_k^2 A 
 +\varepsilon q A& = 0 \, , \nonumber\\
 \ddot{B} + 2\mu \dot{B} - \omega \dot{A} -\omega\mu A -{\Delta\over2}\omega B 
-{\varepsilon\over2}\omega_k^2 B 
 +\varepsilon{q}B& = 0 \,  .\label{eq:v6}
\end{eqnarray}

   Let us initially study the case when the noise is neglected. Since 
$q=0$, by following the Bogoliubov method we can assume that $\dot{A}$,  
$\dot{B}$, $\ddot{A}$ and $\ddot{B}$ are of second order in $\varepsilon$.
The resulting system is  (recalling that $\omega = 2\omega_k +\Delta$)
 \begin{equation}
2\mu A +B\Delta+{\varepsilon\over2}\omega_k B=0 \,  ,\quad 
2\mu B -A\Delta+{\varepsilon\over2}\omega_k A = 0 \,  ,\label{eq:v7}
\end{equation}
which yields  
\begin{equation}
\mu^2 ={\omega_k^2\over4}\left({\varepsilon^2\over4}
-{\Delta^2\over\omega_k^2}\right)\, ,\quad \quad A \sim B =const. 
\label{eq:v8}
\end{equation}
   This result means that as long as $\mu$ is real ($-\varepsilon\omega_k < {2\Delta}<\varepsilon\omega_k$) and for $q=0$ the $\chi_k$ field grows 
exponentially in time.   

{F}rom the above analysis a simple but very important result follows. In 
the narrow resonance regime, the resonance is not destroyed by noise which is 
small compared to the amplitude of the oscillatory inflaton field 
$p(\omega t)$. Specifically, if $g$ (see (\ref{noise2})) is small, say $g= \it{O} (\varepsilon)$, then the last term in both of the equations 
(\ref{eq:v6}) can be neglected and the result of Eqs. (\ref{eq:v8}) follows.

On the other hand, when $q$ is not negligible, then the derivatives $\dot{A}$, $\dot{B}$, $\ddot{A}$ and $\ddot{B}$  must be of first order in $\varepsilon$ and not $O(\varepsilon^2)$ as we assumed above. This can be seen as follows. Solving equations (\ref{eq:v6}) as before under the assumption that the derivatives are negligible leads to a solution for $A$ and $B$ which depends on time (via the noise $q$), and whose derivatives are thus proportional to $\varepsilon \Gamma$, where $\Gamma$ is the rate of change of the noise. This demonstrates that the derivatives are not negligible. 

  In order to circumvent this difficulty we differentiate once equations
(\ref{eq:v6}) and neglect all terms of second and higher order in 
$\varepsilon$ to get the following
\begin{eqnarray} 
A^{^{^{\hskip -.24cm \mbox{\small .\hspace{-.02cm}.\hspace{-.02cm}.}}}}
  +\omega \ddot{B}+\varepsilon \dot{q} A& = 0 \, , \nonumber\\
 B^{^{^{\hskip -.27cm \mbox{\small .\hspace{-.02cm}.\hspace{-.02cm}.}}}}
- \omega \ddot{A} +\varepsilon\dot{q} B& = 0 \,  .\label{eq:v9}
\end{eqnarray}

    By integrating the above system we get (to first order in $\varepsilon$)
\begin{eqnarray} 
\ddot{A } +\omega \dot{B}+\varepsilon q A& = C_1\, , \nonumber\\
 \ddot{B} - \omega \dot{A} +\varepsilon q B& = C_2\,  ,\label{eq:v10}
\end{eqnarray}
where $C_1$ and $C_2$ are both constants of order of $\varepsilon$.

  The next step is to check the compatibility between (\ref{eq:v10}) and 
(\ref{eq:v6}).  Inserting (\ref{eq:v10}) into (\ref{eq:v6}) and neglecting 
terms of order $\varepsilon^2$, we obtain a matrix equation for the vector 
$(A, B)$ which has a solution only if the determinant of the coefficient 
matrix vanishes. This condition leads to the same value of the Floquet 
exponent $\mu$ as obtained in the absence of noise (cf. Eq. ({\ref{eq:v8})). 
Since $q(t)$ is a general time dependent function, a solution exists only 
if $C_1 = C_2 = 0$, and only if the terms $\omega \dot{B}$ and $\omega\dot{A}$ 
in (\ref{eq:v10}) are negligible compared to the other terms. Any other
choice for $C_1$ and $C_2$ implies $A=const.$, $B=const.$, which is only the
case for $g= O(\varepsilon)$ as we showed above.
 
  Moreover, using (\ref{eq:v9}) and (\ref{eq:v10}) it is seen that 
conditions $|\omega \ddot{B}| \ll|\varepsilon {\dot q} A |$ and 
$|\omega\ddot{A}|\ll|\varepsilon {\dot q} B|$  
are equivalent to $\omega \ll {|\dot{q}|} / {|q|}\sim \Gamma$,
where $\Gamma$ is the characteristic rate of the noise. In this 
situation, Equations (\ref{eq:v10}) reduce to $\ddot{A} +\varepsilon q A =0\,$,
$\ddot{B} +\varepsilon q B =0 \,$, whose solutions to first order in
$\varepsilon$ can be written as $A\, , B\propto e^{-\gamma(t)}$, where
$\gamma(t) = \varepsilon\int_o^t \Psi(\tau)d\tau$ with $\dot{\Psi} = q
\equiv g\omega^2 n(t)$ (see Eq. (\ref{noise2})).
  
In this case, parametric resonance can be destroyed by possible
exponential decay of $A$ and $B$. For assume that $t$ is sufficiently big 
and that the noise $n(t)$ is a random walk, so that we can define
the average of $n$ in the time interval from $0$ to $t$ as 
$\bar{n}= {1\over t}\int_o^t n(\tau)d\tau\sim 1/(t\Gamma)^{1/2}$, 
which yields $\gamma(t) \sim g\varepsilon\omega^2(t^3/\Gamma)^{1/2}$.
This leads  to $A$, $B\propto e^{-\varepsilon g\omega^2 \bar{n} t^{3/2} \Gamma^{-1/2}}$,
 which implies that (see Eqs. (\ref{eq:v5})) $a(t)$ and $b(t)$ 
vary exponentially with $\mu t -\gamma(t)$. Since $\gamma(t)$ grows faster 
than $t$, for sufficiently large times and for $\bar{n}>0$ it follows
$\mu t -\gamma(t) <0$. To be precise, let us assume that $\gamma(t)$ is 
positive and define the time $t_e$ such that $\mu t_e - \gamma(t_e)=0$. This
furnishes $t_e \simeq \Gamma/g^2\omega^2$.
Therefore, for $t >t_e$, $a(t)$ and $b(t)$ decay exponentially with time, 
destroying the resonance. However, for 
$\bar{n} < 0$  the resonance is enhanced by the noise.

Notice that $\varepsilon\omega$ is of the same order as $\mu$ so that the noise
is  effective for times $t > t_e$, where $t_e$ is such that $g^2 \omega t_e = \Gamma/\omega \gg 1$. For $g$ sufficiently small, $t_e$  can
be bigger than the whole reheating time, implying no effect on the
exponential growth of the resonant field $\chi$.   This result is consistent with the results of the previous sections and also with the numerical 
studies reported in Sect. VII (see also section VI). Based on the solution without noise, the reheating time can be taken to be of the order $\mu^{-1}$. Thus, a sufficient condition for noise not to prevent the onset of resonance is
\begin{equation} \label{bogores}
g^2 \ll {{\mu} \over {\omega}}{{\Gamma} \over {\omega}} \, .
\end{equation}
Since $\Gamma > \mu$, this condition is less restrictive than the result (\ref{eq:a18}) obtained in the previous section. Note, in particular, that the larger $\Gamma$ is, the less sensitive the resonance is to the effects of the noise.

  Concerning the opposite limit $\omega \gg |\dot{q}|/|q|$ to which the above analysis is not applicable, there is however an alternate procedure to 
study the problem. In such a case, the periodic function oscillates many times
during a characteristic time step of the noise $\Gamma^{-1}$. The noise 
can then be thought as a slowly varying function when compared to $p(\omega t)$.
Therefore, $q(t)$ is ``almost" constant within a time interval of the order of $\Gamma^{-1}$ and can be considered as a perturbation
on the frequency $\omega_k^2$. The analysis via Bogoliubov method can be 
repeated for each of the $N\equiv t\Gamma$ time intervals, but with a different 
frequency squared $\omega_k^2 + \varepsilon q_j$ for each interval. Here $q_j$ 
can be chosen as the extremum value of the noise within the $j$-th 
time interval. The  Floquet exponent is then $\displaystyle{\mu^2=
{1\over4}\left({\varepsilon^2\omega_k^2\over4}-\left(\Delta +\varepsilon 
{q_j\over\omega_k}\right)^2\right)}$.  We see that the resonance is preserved provided that the inequality $-\varepsilon\omega_k < 2(\Delta +\varepsilon q_j/\omega_k)< 
\varepsilon \omega_k$ holds for all time intervals.  The average Floquet
exponent is obtained by replacing $q_j$ in the above expression by its
mean value over all reheating time $\bar{q}\sim g\omega_k^2/(t\Gamma)^{1/2}$.
Notice that (for fixed $\omega_k$) $|\bar{q}|$, and then the effect of the noise, decreases with $\Gamma$ and $t$.  This is also verified 
numerically (see Sect. VII) and it holds approximately even in the case
 $\omega_k \simeq \omega/2 \ll|\dot{q}|/|q|$.

\section{An Estimate by Means of the Born Approximation}             

The two previous methods give (estimates for) lower bounds on the 
amplitude of the noise in the inflaton field for which we can prove 
that parametric resonance persists. However, it is to be expected 
that resonance persists for substantially larger amplitudes. In 
this section, we adopt a perturbative technique to estimate the 
strength of the noise required to change the resonant behavior 
of the modes. 

The starting point is the mode equation (\ref{eq:n1}). We will 
solve this equation in the first order Born approximation. We 
write the solution $\chi$ (dropping the mode index $k$) as
\begin{equation}
   \chi \, = \, \chi_h + \chi_q \, , \label{eq:n31}
\end{equation}
where $\chi_h$ is the solution without noise satisfying the 
given initial conditions, and $\chi_q$ is the contribution 
of the noise to $\chi$ computed to first order in the Born 
approximation, i.e. satisfying the equation
\begin{equation}
   {\ddot \chi_q} + (\omega_k^2 + \varepsilon p(t)) \chi_q \, 
      = \, \varepsilon q(t) \chi_h(t) 
\label{eq:n32}
\end{equation}
and with vanishing initial data. If we introduce a noise 
coefficient $g$ as in (\ref{noise2}), then our approximation 
corresponds to first order perturbation theory in $g$. 

{}For the time dependence of (\ref{noise}), i.e. 
$p(t) \sim {\rm cos}(\omega t)$, the ``homogeneous" 
solution $\chi_h$ can be (to first order in $\varepsilon$) 
written as
\begin{equation}
   \chi_h(t) \, = \, c_1 e^{\mu t} {\rm cos}({\omega \over 2}t + \varphi_1)
   \, + \, c_2 e^{- \mu t} {\rm cos}({\omega \over 2}t + \varphi_2) \, ,
\label{eq:n33}
\end{equation}
where $c_1$ and $c_2$ are the amplitudes of the two fundamental 
solutions of the homogeneous equation (denoted $\chi_1(t)$ and 
$\chi_2(t)$), and $\varphi_1$ and $\varphi_2$ are phases.

By means of the Greens function method, the solution of 
(\ref{eq:n32}) takes the form
\begin{equation}
   \chi_q(t) \, = \, \chi_1(t) \int_{t_i}^t dt^{\prime} 
    W(t^{\prime})^{-1} \chi_2(t^{\prime}) s(t^{\prime}) \, 
    - \, \chi_2(t) \int_{t_i}^t dt^{\prime} W(t^{\prime})^{-1} 
    \chi_1(t^{\prime}) s(t^{\prime}) \, ,
\label{eq:n34}
\end{equation}
with a ``source" term
\begin{equation}
   s(t) \, = \, \varepsilon q(t) \chi_h(t)
\label{source}
\end{equation}
and with the Wronskian
\begin{equation}
    W(t) \, = \, {\dot \chi_1(t)} \chi_2(t) - {\dot \chi_2(t)} \chi_1(t) \, 
    \simeq \, {\omega \over 2} {\rm sin}(2\varphi)
\label{Wronskian}
\end{equation}
(making use of $\varphi = \varphi_1 = - \varphi_2$). As mentioned 
in Section 3, the Wronskian is time-independent.

Inserting the expressions for the source (\ref{source}) and for 
the mode functions $\chi_1$ and $\chi_2$ (see (\ref{eq:n33})), 
and neglecting the contribution of the decaying mode $\chi_2$ in 
the source, we obtain
\begin{equation}
   \chi_q(t) \, = \, c_1 \varepsilon W^{-1} e^{\mu t} 
     {\rm cos}({\omega \over 2}t + \varphi) I_1(t) \, 
     - \, c_1 \varepsilon W^{-1} e^{- \mu t} {\rm cos}
       ({\omega \over 2}t - \varphi) I_2(t) \, ,
\label{eq:n35}
\end{equation}
where the integrals $I_1$ and $I_2$ are
\begin{equation}
     I_1(t) \, = \, \int_{t_i}^t dt^{\prime} q(t^{\prime}) 
     {\rm cos}({\omega \over 2}t^{\prime} + \varphi)
     {\rm cos}({\omega \over 2}t^{\prime} + \varphi)
\label{I1}
\end{equation}
and
\begin{equation}
     I_2(t) \, = \, \int_{t_i}^t dt^{\prime} e^{2 \mu t^{\prime}} 
     q(t^{\prime}) {\rm cos}^2({\omega \over 2}t^{\prime} + \varphi) \, .
\label{I2}
\end{equation}

In estimating the magnitudes of the integrals $I_1$ and $I_2$ we 
will for the first time make use of our assumption of random noise. 
To be more specific, we will model the noise function $n(t)$ (see 
(\ref{noise2})) as a random walk with unit amplitude and step 
length $\Gamma^{-1}$ (note that this is a time interval!). By 
inserting (\ref{noise2}) into (\ref{I1}) and using the standard 
formula for the ``radius" of a random walk in terms of the 
individual random step length, we obtain the estimates
\begin{equation}
     I_1 \, \sim \, g \omega^2 (t \Gamma)^{1/2} \Gamma^{-1} 
\label{I1est}
\end{equation}
and
\begin{equation}
     I_2 \, \sim \, g \omega^2 e^{2 \mu t} ({{\Gamma} \over {\mu}})^{1/2} 
     \Gamma^{-1}
\label{I2est}
\end{equation}
(where we have set the initial time $t_i = 0$ to simplify the 
notation). We conclude that the contribution of $I_1$ dominates 
in (\ref{eq:n35}) as long as we consider time intervals 
$t \geq \mu^{-1}$. Thus, we get the following estimate for $\chi_q$:
\begin{equation}
     \chi_q(t) \, \sim \, c_1 \varepsilon g \omega^2 W^{-1} e^{\mu t} 
    (t \Gamma)^{1/2} \Gamma^{-1} \, ,
\label{eq:n36}
\end{equation}
which must be compared with the amplitude of the ``homogeneous" mode
\begin{equation}
    \chi_h(t) \, \sim \, c_1 e^{\mu t} \, .
\label{eq:n37}
\end{equation}

By demanding that the contribution of $\chi_q$ be smaller 
than $\chi_h$ over a typical time interval for resonance 
(i.e. for $t \sim \mu^{-1}$), and inserting the value 
(\ref{Wronskian}) of the Wronskian, we obtain as the 
condition under which the noise has a negligible effect 
on parametric resonance:
\begin{equation}
     g \, < \, {{2 \Gamma} \over {\omega \varepsilon}} \mu (\mu \Gamma)^{-1/2} 
     {\rm sin}(2 \varphi) \, .
\label{eq:n38}
\end{equation}
To discuss the consequences of this condition, recall the 
expression (\ref{eq:v9}) for $\mu$. The value of $\mu$ is 
maximal in the center of the resonance band and vanishes 
at the band edges. Hence, we conclude from (\ref{eq:n38}) 
that any noise will tend to slightly decrease the width 
of the resonance bands. However, resonance at the central 
value of $k$ is not significantly effected unless
\begin{equation}
     g \, > \, {1 \over 4} \bigl({\Gamma \over \mu}\bigr)^{1/2} 
     {\rm sin}(2 \varphi) \, ,
\label{eq:n39}
\end{equation}
which in general yields a value greater than 1. Note, however, 
that for $g \geq 1$ the Born approximation is an invalid 
perturbative expansion. 

We conclude that even large amplitude noise is unlikely to 
interfere with parametric resonance, and that in fact the 
shorter the time period of the noise, the less likely the 
noise is to influence the resonant modes (this is reminiscent 
of the Riemann-Lebesgue Lemma).   

\section{Results of Numerical Studies}

  At this point we show that the results obtained in the previous
sections based on analytical approximate methods are perfectly consistent 
with the numerical analysis of the problem. We also verify 
that the results of previous sections hold for both the narrow resonance and 
the broad resonance cases, which we studied separately. 

  Following the notation of previous sections, we have chosen $p(\omega t)=
A\cos(2\omega t)$, where $A$ is a constant, and taken the noise
$q(t)=g\,A\,n(t)$ with $g$ being a positive constant and with $n(t)$ 
being a random function of time with characteristic time $\Gamma^{-1}$ and 
amplitude 1. To simplify the analysis we redefine the time variable
to a dimensionless time $t \longrightarrow t=\omega t$ so that  
the evolution equation for $\chi_k$ reads
\begin{equation}
\ddot{\chi} +\left[E^2 + \lambda \cos(2t) +g\lambda n(t)\right]\chi =0\, ,
\label{eq:vii1} \end{equation}
where  $\lambda\equiv A/\omega^2$ and $E^2\equiv \omega_k^2/\omega^2$.

In the narrow resonance regime, the first resonance band 
($ E^2 =1$) is the most important. This is because both the width of the 
$N$-th band and the correspondent  Floquet exponent are $O(\lambda^N)$. 
This implies that
the value of $E^2$ can be shifted (e. g. due to the presence of the noise) 
by an amount comparable to (but smaller than) $\lambda$ without moving 
a resonant mode into a stability band. 
Only if the amplitude of the noise function is large ($g > 1/2$) it is 
possible that the resonance is significantly affected (see Fig. 1).

  To be explicit let us recall the main characteristics which differentiate
the two regimes of parametric resonance analysed here: Narrow resonance 
is defined as the regime where $|\lambda| \ll 1$; $E^2 = N^2 +O(\lambda)$, 
$N =1$, $2$, ...; and Floquet (Lyapunov) exponent $\mu \ll 1$. Broad resonance is characterized  by $|\lambda| >1$; 
$E^2 \geq |\lambda|$ and $\mu \sim O(1)$.    
In the broad resonance regime $\lambda$ can be much bigger than $1$
but the width of the instability may not be of the same order of 
$\lambda$ , so that even a small change in $E^2$ 
(when compared with $E^2$ itself) can be enough to 
shift a mode  out of the resonance band. This implies that even noise  
with $g \ll 1$ can be important. Moreover, since the growth rate 
is large ($\mu =O(1)$), the particular behavior of $\chi(t)$ is much more
sensitive to the functional form of the noise in the broad resonance than in 
the narrow resonance regime (see Fig. 2 and the discussion below).

In our numerical work, equation (\ref{eq:vii1}) was solved using a MATLAB integration routine, with a noise function which was taken to be piecewise constant in intervals of length $\Gamma^{-1}$, and whose amplitude was chosen at random from a uniform distribution on the interval $(-1, 1)$.
The most important numerical results can be summarized as follows (all analyses were done for $g \leq 1$):
\begin{figure}
\begin{center}
   \mbox{\epsfig{figure=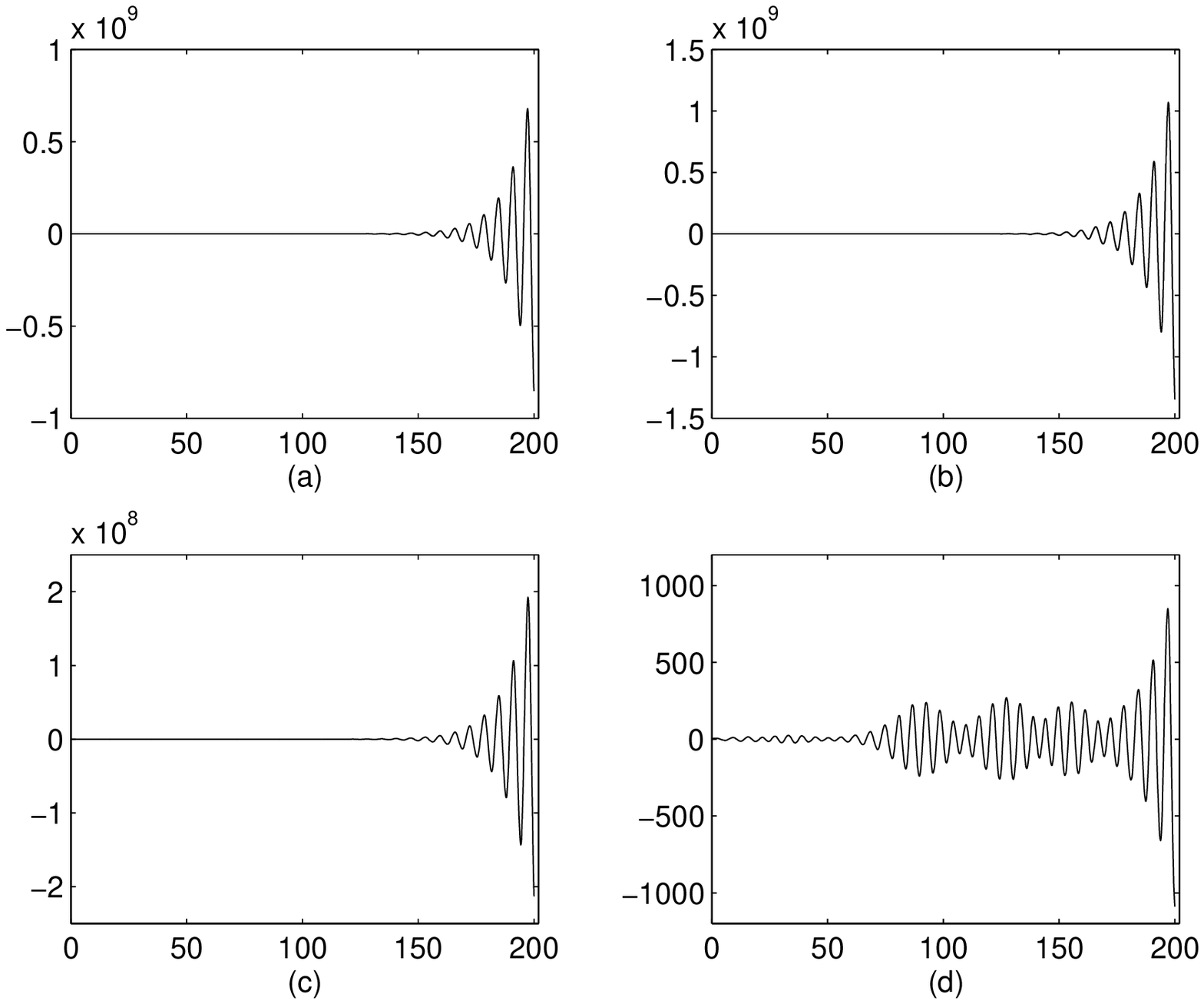,height=4.2in}}
\end{center}
\caption{ Sample of narrow resonance regime results.  $\chi_k(t)$ is plotted
(vertical axis) against dimensionless time (horizontal axis) for a value of 
$k$ in the first resonance band, $E^2 =1$, and for $\lambda= 0.4$.\\
(a) No noise is present, $q(t)=0$.\\
(b) The noise is characterized by $\bar{q} \approx 0.010\, \lambda$, 
$\Gamma = 10$ and $g = 0.8$ (large amplitude and large noise frequency).\\
(c) The noise is characterized by $\bar{q} \approx 0.028\, \lambda$, 
$\Gamma = 1/40$ and $g =0.4$ (small amplitude and small noise frequency).\\
(d) The noise is characterized by  $\bar{q} \approx 0.433\, \lambda$, 
$\Gamma = 1/20$ and $g = 0.9$ (large amplitude and small noise frequency). 
In this case, the mode $k$ is shifted out of the resonance band. This is 
an example for which the noise is important.}
\end{figure}

  (i) The noise has practically no effect when $\Gamma$, the inverse of its
characteristic time, is much bigger than the frequency $\omega$ of the 
periodic driving function $p(\omega t)$, i.e. when $\Gamma \gg \omega$. This is shown in Figures 1(b) and 2(b) where $\Gamma \gg 1$, and it is consistent with 
the results of Sections V and VI (see e.g. Eqs. (\ref{eq:n38}) and (\ref{eq:n39})). On the other hand, if $\Gamma$ is small compared to $\omega$, then the ergodic hypothesis is not satisfied, and the noise may completely alter the efficiency of energy transfer (see e.g. Figures 1(d) and 2(d)).

 (ii) In the case of narrow resonance, the noise is effective only if its
magnitude is of the same order or larger than the amplitude of $p(\omega t)$, i.e. $g > 1/2$. For values of $g$ significantly smaller than $1$ ($g< 0.5$, say) the noise is not effective for ``central" modes in a resonance band.  This result was obtained in Sections III, IV and V and is illustrated by 
Figure 1(c).
  
 (iii) The presence of noise may lead to a much faster exponential growth of
a mode initially in a resonance band. This effect is particularly 
important in the case of broad resonance (Figure 2(d)). 

  (iv) After having analyzed the time evolution of $\chi_k(t)$ for many
different random walks we verified that an important quantity
to be considered is $\bar{q}$, the end point distance of $q(t)$ divided by the number of steps, over the total time
interval of reheating $t$. Note that $\bar{q} =g\lambda\bar{n} \sim 
g\lambda/(t\Gamma)^{1/2}$.  If $\bar{q}$ is
small compared to the amplitude $A$ of $p(\omega t)$, i. e. if
$g/(t\Gamma)^{1/2} \ll 1$, then the resonance is preserved. For some very special random walks in the broad resonance regime (see
Figure 2(c)) there are specific realizations for which the noise decreases the Lyapunov exponent, but when taking the average over many realizations, agreement with the analytical results is restored. 

   In order to understand the above results, notice that
parametric resonance of a given mode of the field 
$\chi$ disappears if the noise is able to keep the mode out of
the resonance band during a sufficiently large time interval (when
compared to the total evolution time). 
\begin{figure}
\centerline{\epsfig{file=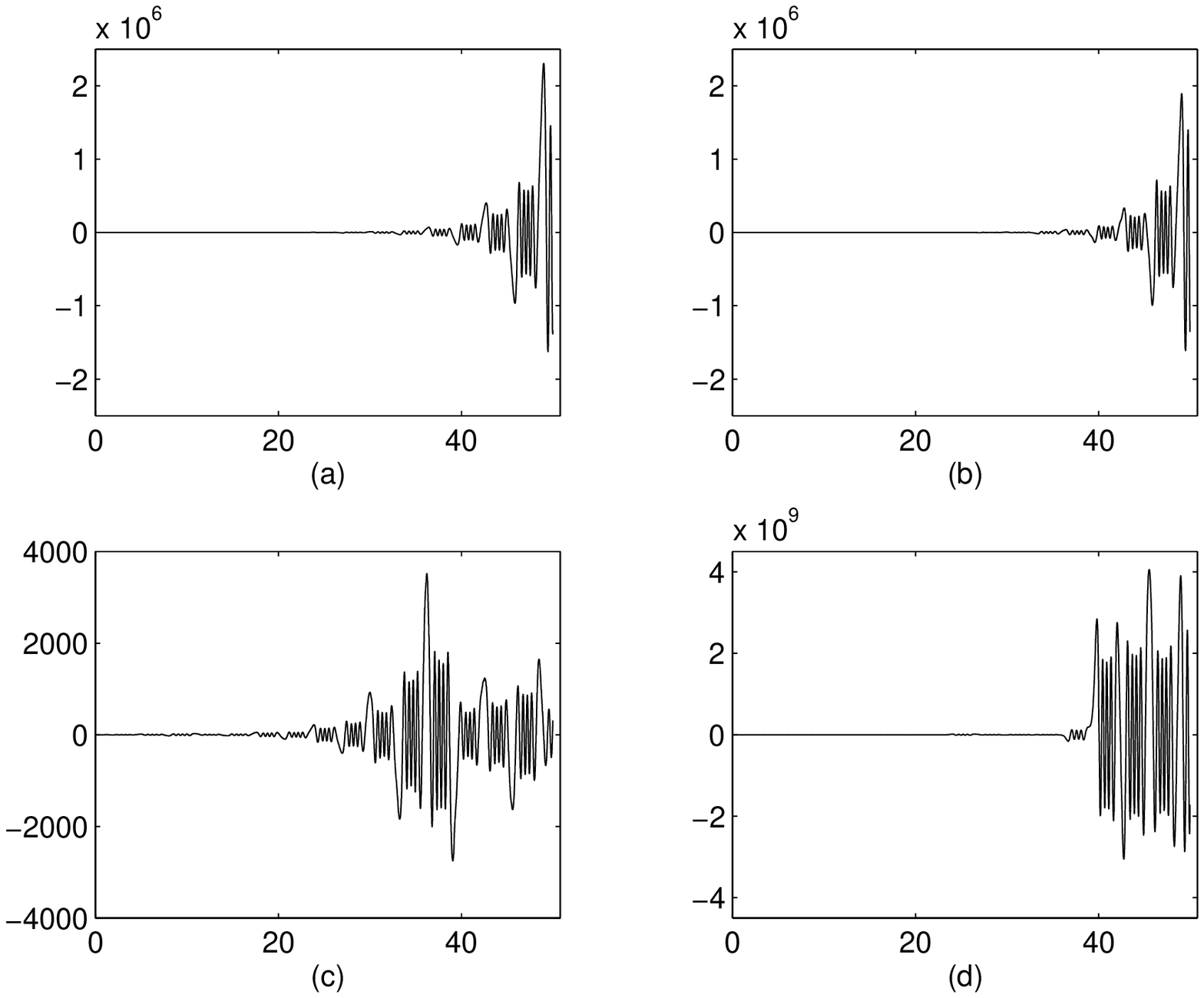,height=4.5in}}
\caption{ Similar plots as in Figure 1, but now for the broad resonance
regime. We consider a mode in the tenth resonance band, $E^2 =100$, and we choose $\lambda= 100$.\\
(a) No noise is present, $q(t)=0$.\\
(b) The noise is given by $\bar{q} \approx -0.004\, \lambda$, $\Gamma=12.5$ and 
$g=0.2$ (i.e. large inverse characteristic time and large noise amplitude).\\
(c) The noise is given by $\bar{q} \approx 0.013\, \lambda$, with the same 
$\Gamma$  and $g$ as in the case (b).\\
(d) The noise is given by $\bar{q} \approx -0.025\, \lambda$, $\Gamma = 0.2$
and $g=0.30$ (small $\Gamma$ and small noise amplitude). In this example the noise has an important effect. It greatly increases the rate of particle production. In this particular case, the noise $n$ takes on
negative values for a few successive time steps, which keeps the mode
in an instability region whose Floquet multiplier is larger than without noise. Thus, even though $\bar{q}$ is small, the overall effect is an increased resonance strength due to the
very fast exponential growth during a short time interval.}
\end{figure} 
  
    To see exactly what this means let us write the band number $N$ in terms 
of the parameters of Eq. (\ref{eq:vii1}).  From the theory of the Mathieu 
equation we have (neglecting the noise) $N_o = |E|$.  According to the 
point (iv) above, the presence of the noise can be approximately taken into 
account by shifting the value of $E^2$ to 
$E^2 + g\lambda\bar{n}= E^2 + \bar{q}/\omega^2$.  Then, by defining $N^2 =N_o^2+\bar{q}/\omega^2$ 
where $N_o$ is the band number when the noise is neglected ($N_o^2 = E^2$) 
we obtain
\begin{equation}
\Delta N \equiv N- N_o = N_o\frac{\bar{q}}{2\omega_k^2}= 
N_o g {\lambda\over E^2} \bar{n} \label{eq:vii2} \, .
\end{equation}

 This result means that for large $N_o$, that is to say for modes initially
in one of the resonance bands with large band number, even a very small mean
noise (when compared to the energy of the mode $\omega_k^2$) can shift a 
given mode through several bands and occasionally into a stability band.
However, since the same noise is assumed to act independently upon all
the modes, it is easy to see that some of the modes initially in a stability
band can be shifted to an instability band. 
 This effect is important in the broad resonance regime, where the resonance 
occurs for $E^2 \sim \lambda \gg 1 $.  For instance, in the case of Fig. 2 
we have $\lambda/E^2=1$, so that 
$\Delta N = 5 g \bar{n}$.  Then, even for $g\approx 0.1$ there are 
special samples of the noise that have a significant effect on the 
resonance of the boson field (a small effect of the noise means 
$\Delta N \ll 1$) . In particular, for Figs. 2(b) and 2(c) we find
$\Delta N \simeq -0.02$ and $\Delta N \simeq 0.08$, respectively. 

 On the other hand, for small $N_o$ in principle the mean noise could be 
of the order of $\omega_k^2$ without jumping
to the next resonance band.  However, this corresponds to the narrow 
resonance regime $\lambda \equiv \varepsilon \ll 1$,  where the width of the 
instability bands are very narrow and  the mode can easily be shifted
to a stability band.  In fact, for a  mode $k$ initially in the center of 
the first band the  mean noise $\bar{q}$ must satisfy $|\bar{q}|/\omega_k^2 
< \varepsilon/2$ in order for not to take the mode out of the first 
resonance band. 
 Notice that $\varepsilon$ is by definition the ratio
$A/\omega_k^2$ so that the (narrow) resonance is preserved as long as 
$|\bar{q}|$ is smaller than one half of the amplitude $\lambda$
of the periodic driving function $p(\omega t) = \lambda \cos(2 t)$.  In the 
case of Fig. 1 we have $\lambda /E^2 =0.4$, so that $\Delta N = 0.4 g \bar{n}$,
 which implies a small effect of the noise even for $g$ of the order 
of $1$. For instance, in the case of Figs. 1(c) and 1(d) we have 
respectively $\Delta N \simeq 0.01$ and $\Delta N \simeq 0.08$.

According to the general analysis of Sect. II, ergodic noise always 
increases the rate of particle production. In the numerical work, however,
we find examples where over a small time interval the rate of particle 
production is decreased as a consequence of the noise.  This is 
manifest in the case of small 
$\Gamma$ (Figures 1(c), 1(d) and 2(d)). This apparent contradiction disappears once we realize that for small $\Gamma$ and for a finite reheating interval, 
we are in fact not 
taking the average of $q(\kappa,t)$ over many realizations in the
sample space $\Omega$, but instead choosing only a few special noise 
functions for each calculation. By the ergodic hypothesis used in 
Section II, we should expect that numerical results to agree with the exact
results only if the reheating time $t$ is very big compared to
the characteristic time $T \equiv \Gamma^{-1}$ of the noise.  
(The time $t$ must be very big, compared to $T$, to ensure that the noise explores all possible realizations in the ``noise space".) This condition 
is grossly violated for the parameters of Figs. 1c and 1d. There are also
cases (see Figure 2(c)) when $\Gamma$ is large but for certain realization 
of the noise the Lyapunov exponent decreases. In all these cases, however, the 
mean Lyapunov number over many realizations is larger than the Lyapunov 
exponent in the absence of noise. It would be of interest to further study 
the dispersion of the Lyapunov exponents for identical values of the physical parameters.

\section{Conclusions}
 
We have studied some effects of noise on parametric resonance. 
Specifically, we considered spatially homogeneous noise in the 
time dependent mass responsible for the parametric resonance 
instability. Assuming that the noise is ergodic we showed that the presence
of the noise leads to strict increase in the rate of particle production. 
We demonstrated also that the resonance is rather 
insensitive to the presence of small noise.   
Under the assumption 
that the time dependence of the noise can be modeled as a 
random walk with a characteristic step length, we derived 
estimates for the amplitude of the noise for which it can 
be shown that the resonance persists. We demonstrated that 
even if the dimensionless amplitude of the noise $g$ is 
of the order 1, resonance is not affected provided the 
time step $\Gamma^{-1}$ of the noise is sufficiently 
short. In a subsequent letter$^{\cite{ZMCB2}}$ we will 
extend these results to the more interesting case of 
spatially inhomogeneous noise.

\section*{Appendix}

This last section gives an outline of the mathematical results
that are used in this work. Complete proofs appear in the 
literature citations. 

\medskip\noindent
{\bf Proposition 1: \ }{\it 
When $q(t;\kappa)$ is given by a translation invariant 
ergodic measure $dP(\kappa)$ on $C({\mathbf R})$, the limit 
exists,
\[
   \overline{\mu} = \lim_{N \to \infty} {1 \over NT} 
      E( \log \|\Pi_{j=1}^N \Phi_q(jT, (j-1)T) \| ) ~.
\]
Furthermore, for almost every realization $\kappa$ (with respect to
the probability measure $dP(\kappa)$) the individual limits exist,
\[
   \mu(q) = \lim_{N \to \infty} {1 \over NT} 
      \log \|\Pi_{j=1}^N \Phi_q(jT, (j-1)T) \| ~,
\]
and they are equal to $\overline{\mu}$.
}\medskip

\noindent
The proof of the first statement follows from an argument involving the
sequence $e(n) = E( \log \|\Pi_{j=1}^n \Phi_q(jT, (j-1)T) \| )$ and its
subadditive property, $e(n+m) \leq e(n) + e(m)$. The proof of the second
statement follows from the subadditive ergodic theorem, \cite{pastur}. 

\medskip\noindent
{\bf Proposition 2: \ }{\it
The generalized Floquet exponent (Lyapunov exponent) $\mu(q)$ is
continuous with respect to $q$ in the topology on $C({\mathbf R})$
of uniform convergence on compact sets. 
}\medskip

This result and its proof may be found in \cite{carmona} and 
\cite{kotani}; it follows from Sturm - Liouville theory and the 
nesting property of the Weyl limit circles.

The third qualitative result has to do with a monotonicity property
of the generalized Floquet exponent in ergodic systems. Consider the
probability space $\Omega = C({\mathbf R})$ with two ergodic invariant 
measures $dP_1$ and $dP_2$. By Proposition 1 the two associated 
generalized Floquet exponents $\mu_1$ and $\mu_2$ are constant 
almost everywhere on the support of their respective measures. The 
following result states that the class of problems (\ref{matrixeq}) with
positive generalized Floquet exponent is nondecreasing with respect 
to the support of the measure.

\medskip\noindent
{\bf Theorem 3: \ }{\it
(S.~Kotani \cite{kotani}) \ If ${\mathrm supp} \, (dP_1) \subseteq 
{\mathrm supp} \, (dP_2)$ and $\mu_1 > 0$, then $\mu_2 > 0$.
}\medskip

In our case $dP_1$ is supported on the periodic function $p(t)$ and its
translates, and $dP_2$ will be taken to be the description of the statistics 
of the realizations $\{ p(t) + q(t;\kappa) \} \subseteq C({\mathbf R})$. 
If $q(t) = 0$ is a possible realization in the support of the probability 
measure $dP(\kappa)$, then the support of $dP_2$ contains the support of 
$dP_1$. We are therefore in the situation described in Theorem~3.

The final mathematical result of this article is also the central one 
to our argument. Consider a probability distribution $dA$ on the matrices
$\Psi \in {\mathrm SL}(2, {\mathrm R})$ (in fact the result applies more 
generally to ${\mathrm SL}(n, {\mathrm R})$). Let $G_A$ be the smallest 
subgroup of ${\mathrm SL}(2, {\mathrm R})$ containing the support of $dA$.

\medskip\noindent
{\bf Theorem 4: \ }{\it
(Furstenberg, \cite{pastur}) \ Suppose that $G_A$ is not compact, and that 
the action of $G_A$ on the set of lines in ${\mathrm R^2}$ has no invariant 
measure. Then for almost all independent random sequences 
$\{ \Psi_j\}_{j=1}^\infty \subseteq {\mathrm SL}(2, {\mathrm R})$ with
common distribution $dA$,
\[
   \lim_{N\to\infty} {1 \over N} 
      \log (\| \Pi_{j=1}^\infty \Psi_j \|) = \lambda > 0 ~.
\]
Furthermore, for given $v_1, v_2 \in {\mathrm R^2}$, then
\[
    \lim_{N\to\infty} {1 \over N} 
      \log ( \langle v_1, \Pi_{j=1}^\infty \Psi_j v_2 \rangle ) = \lambda 
\] 
for almost every sequence $\{ \Psi_j\}_{j=1}^\infty$. 
}\medskip

In our setting, realizations $\{ q(t;\kappa) : 0 \leq t < T \}$ give rise
to transfer matrices $\Psi_q(T,0) \in {\mathrm SL}(2, {\mathrm R})$, and
the probability measure $dP$ on $C({\mathbf R})$ induces a measure $dA$
on ${\mathrm SL}(2, {\mathrm R})$. We will fulfill the hypotheses of the 
theorem of Furstenberg for $dA$ by demonstrating that under our conditions 
on ${\mathrm supp} (dP)$, then $G_A = {\mathrm SL}(2, {\mathrm R})$. This 
will  follow if we show that the set of random transfer matrices 
$\{ \Psi_q(T,0) : q \in {\mathrm supp} (dP)\}$ contains a small 
neighborhood of the identity in ${\mathrm SL}(2, {\mathrm R})$, for 
this must be contained in the support of the induced measure $dA$. 
In order to analyse this set, observe that the sequence of successive
approximations $\{ \Psi_j(t,0)\}_{j=1}^\infty$ gives the Taylor 
polynomials of the fundamental solution $\Psi(t,0;q) = \Psi_q(t,0)$ with 
respect to $q \in C(0,T)$ about $q(t) = 0$. Thus the derivative of
the transfer matrix $\Psi(T,0;q)$ with respect to $q(t)$ at zero is
given by the first term
\begin{equation}
   \delta_q \Psi(T,0;0) \cdot r(t) = 
   \int_0^T \ dt \ \ r(t) \Phi_0^{-1}(t) 
         \left( \begin{array}{cc} 
            0  & 0 \\  
            -1 & 0   
         \end{array} \right) \Phi_0(t)  ~.
\end{equation}
{}From (\ref{matrixsol}) and the fact that ${\mathrm det}(\Phi_0(t)) = 1$, 
this is
\begin{eqnarray}\label{d_Psi}
   \delta_q \Psi(T,0;0) \cdot r(t) & = & 
   \int_0^T \ dt \ \ r(t) 
        \left( \begin{array}{cc} 
            \dot\phi_2 & -\phi_2 \\  
            -\dot\phi_1 & \phi_1  
         \end{array} \right) 
         \left( \begin{array}{cc} 
            0  & 0 \\  
            -1 & 0   
         \end{array} \right) 
         \left( \begin{array}{cc} 
            \phi_1  & \phi_2 \\  
            \dot\phi_1 & \dot\phi_2   
         \end{array} \right)                       \\
     & = & \int_0^T \ dt \ \ r(t) 
       \left( \begin{array}{cc} 
            \phi_1(t)\phi_2(t)  & \phi_2^2(t) \\  
            -\phi_1^2(t) & -\phi_1(t)\phi_2(t)  
         \end{array} \right) ~~.                      
\end{eqnarray}
For $r(t) \in C(0,T)$ the expression $\delta_q\Psi(T,0;0) \cdot r$
is in the Lie algebra ${\mathrm sl}(2, {\mathrm R})$. Taking 
$r(t) \in {\mathrm supp} (dP)$ in a small neighborhood of zero, if
$\delta_q\Psi(T,0;0) \cdot r$ spans ${\mathrm sl}(2, {\mathrm R})$,
then by the implicit function theorem the solutions $\Psi(T,0;q)$ will
indeed fill a neighborhood of $I \in {\mathrm SL}(2, {\mathrm R})$.
{}From the structure of (\ref{d_Psi}) the rank of 
$\delta_q\Psi(T,0;0) \cdot r$ is at most three, spanned by 
$\{ \phi_1^2(t), \phi_1(t)\phi_2(t),  \phi_2^2(t)\}$, and the question 
is whether these three components are in all cases linearly independent
in $C(0,T)$. 

By direct calculation one verifies that products 
$\psi(t)= \phi_j(t)\phi_k(t)$ of solutions of equation (\ref{eq:n1}) 
satisfy themselves a third order differential equation
\begin{equation}\label{thirdorder}
   {d^3 \over dt^3} \psi = -2({d \over dt} (\omega^2 + p(t)) 
       + (\omega^2 + p(t)){d \over dt})\psi ~.
\end{equation}
At $t=0$ the Wronskian for (\ref{thirdorder}) is
\begin{equation}
   \Delta(t)\Bigl|_{t=0} = \left( \begin{array}{ccc} 
            \phi_1^2 & \phi_1\phi_2 &  \phi_2^2 \\  
            \dot{(\phi_1^2)}  & \dot{(\phi_1\phi_2)} & \dot{(\phi_2^2)}    \\
            \ddot{(\phi_1^2)}  & \ddot{(\phi_1\phi_2)} & \ddot{(\phi_2^2)} \\
         \end{array} \right)
         = \left( \begin{array}{ccc}
             1 & 0 & 0 \\
             0 & 1 & 0 \\
             -2(\omega^2 + p(t)) & 0 & \ \ \ \ 2 \ \ \ \    \\ 
           \end{array} \right) ~,
\end{equation}
therefore $\det(\Delta(t)) = 2$ for all $t$, and the three functions
$\phi_1^2(t), \phi_1(t)\phi_2(t),  \phi_2^2(t)$ form a linearly 
independent set. 

Suppose by way of contradiction that the rank of 
$\delta_q\Psi(T,0;0) \cdot r$ is less than three, so that there is a 
linear relationship 
\[
   c_1 \phi_1^2 + c_2 \phi_1(t)\phi_2(t) + c_3 \phi_2^2(t) = 0 ~.
\]
Taking two derivatives, this implies that $(c_1,c_2,c_3)^t$ is a 
null vector for $\Delta(t)$, contradicting the above assertion of 
independence. 

It is therefore only necessary that 
${\mathrm span}(\phi_1^2(t), \phi_1(t)\phi_2(t),  \phi_2^2(t)) 
\cap {\mathrm supp} (dP)$ contain a small neighborhood of the origin 
in order that
 
${\mathrm rank} \{ \delta_q\Psi(T,0;0) \cdot r :
r(t) \in {\mathrm supp} (dP) \cap B_\epsilon(0) \} = 3$. 

This is surely
satisfied for any periodic potential $p(t)$ if 
$B_\epsilon(0) \subseteq {\mathrm supp} (dP)$, which is our hypothesis.

\section*{Acknowledgments}

This work is partially supported by CNPq and FAPESP (Brazilian Research 
Agencies), and by the US Department of Energy under contract DE-FG0291ER40688,
 Task A .

\end{document}